\documentclass[12pt]{article}
\usepackage{amsmath,amssymb,bm,graphicx,epsfig}
\usepackage{lineno,booktabs,url}
\include{epsf} 

\newcommand\as{\alpha_S}

\def\tL{{\tilde L}}

\def\beqn{\begin{eqnarray}} 
\def\eeqn{\end{eqnarray}} 
\def\beq{\begin{equation}} 
\def\eeq{\end{equation}}

\begin{document}
\begin{titlepage}
\begin{flushright}
\end{flushright}

\renewcommand{\thefootnote}{\fnsymbol{footnote}}
\vspace*{2.5cm}

\begin{center}
{\Large \bf 
  Drell--Yan lepton-pair production: 
  \\[0.1cm]
  $\bf q_T$ resummation at N$\bf ^3$LL  accuracy \\[0.1cm]
  and fiducial cross sections at
  N$\bf ^3$LO\\[0.1cm]
}
\end{center}

\par \vspace{2mm}
\begin{center}
  {\bf Stefano Camarda${}^{(a)}$},  {\bf Leandro Cieri${}^{(b)}$}
  and {\bf Giancarlo Ferrera${}^{(c)}$}\\

\vspace{5mm}

${}^{(a)}$ 
CERN, CH-1211 Geneva, Switzerland\\\vspace{1mm}

${}^{(b)}$ 
INFN, Sezione di Firenze,
I-50019 Sesto Fiorentino, Florence, Italy\\\vspace{1mm}

${}^{(c)}$ 
Dipartimento di Fisica, Universit\`a di Milano and\\ INFN, Sezione di Milano,
I-20133 Milan, Italy\\\vspace{1mm}

\end{center}

\vspace{1.5cm}

\par \vspace{2mm}
\begin{center} {\large \bf Abstract} \end{center}
\begin{quote}
\pretolerance 10000
We present   high-accuracy QCD predictions for the transverse-momentum ($q_T$)
distribution  and fiducial cross sections
of Drell--Yan lepton pairs produced in hadronic collisions.
At small value of $q_T$
we resum to all perturbative orders the
logarithmically enhanced contributions
up to next-to-next-to-next-to-leading logarithmic (N$^3$LL) accuracy,
including all the next-to-next-to-next-to-leading order (N$^3$LO)
(i.e.\ $\mathcal{O}(\alpha_S^3)$) terms.
Our resummed calculation has been implemented in the public
numerical program {\ttfamily DYTurbo}, which produces fast and
precise predictions with the full dependence on
the final-state leptons kinematics.
We consistently combine our resummed results with the
known $\mathcal{O}(\alpha_S^3)$ fixed-order predictions at large values of $q_T$
thus obtaining full N$^3$LO accuracy also for fiducial cross sections.
We show numerical results at LHC energies discussing the reduction
of the perturbative uncertainty with respect to lower-order calculations.

\end{quote}

\vspace*{\fill}
\vspace*{2.5cm}

\begin{flushleft}
March 2021
\end{flushleft}
\end{titlepage}

\setcounter{footnote}{0}
\renewcommand{\thefootnote}{\fnsymbol{footnote}}


After the successful operation of the first two runs of the Large Hadron Collider (LHC) at CERN
and the discovery of the long sought Higgs boson, a major task of the high-energy physics community
has become a direct investigation of the electroweak symmetry breaking mechanism.
In the absence of clear direct signals of new physics phenomena, precision studies
give us a unique opportunity to search for possible deviations from Standard Model (SM) predictions.
In this scenario it is clear that theoretical predictions for SM cross sections and
associated distributions at an unprecedented level of accuracy are indispensable to fully exploit the
discovery potential provided by the collected and forthcoming collider data. 

The electroweak (EW) vector boson 
production,  through the Drell--Yan (DY) mechanism\,\cite{Drell:1970wh,Christenson:1970um},
is the most ``classical'' hard-scattering process in hadronic collisions.
The large production rates and clear experimental signatures
make this processes important for detector calibration, as luminosity monitor and to probe underlying event.
Moreover it plays a fundamental role in the contest of SM precision studies\,\cite{Chatrchyan:2011ya,Aad:2015uau,Khachatryan:2016pev,Aaboud:2016btc,Camarda:2016twt} and for 
searches of physics signals beyond the SM\,\cite{Khachatryan:2016zqb,Khachatryan:2016jww,Aaboud:2016cth,Aaboud:2016zkn}.
It is thus essential to provide accurate
theoretical predictions, through detailed computations of
the higher-order radiative corrections in QCD and in the EW theory,
for vector boson production cross sections and related kinematical
distributions.
Among the various kinematical distributions the
vector boson transverse-momentum ($q_T$) spectrum plays a special role.
Precise knowledge of the $Z$ boson $q_T$
distribution gives important information on the $W$ 
boson spectrum which in turn
directly affects the measurement of the $W$ boson mass\,\cite{Abazov:2012bv,Aaltonen:2012bp,Aaboud:2017svj}.

The next-to-next-to-leading 
order (NNLO) corrections in the QCD coupling $\alpha_S$ have been computed for the total cross section\,\cite{Hamberg:1990np,Harlander:2002wh}, the rapidity distribution\,\cite{Anastasiou:2003ds}
and at fully differential level including the leptonic decay of the vector boson\,\cite{Melnikov:2006di,Melnikov:2006kv,Catani:2009sm,Catani:2010en}. More
recently next-to-next-to-next-to-leading
order (N$^3$LO) QCD calculations of the total cross section have been performed in Refs.\,\cite{Duhr:2020seh,Duhr:2020sdp}.
The next-to-leading order (NLO) EW corrections, the mixed QCD-EW and QCD-QED corrections have also been
computed\,\cite{Dittmaier:2001ay
,Baur:2004ig,Zykunov:2006yb,Arbuzov:2005dd,CarloniCalame:2006zq,Baur:2001ze,Zykunov:2005tc,CarloniCalame:2007cd,Arbuzov:2007db,Kotikov:2007vr,Kilgore:2011pa,Dittmaier:2014qza,Bonciani:2016wya,Bonciani:2019nuy,Bonciani:2020tvf,Cieri:2018sfk,deFlorian:2018wcj,Delto:2019ewv,Cieri:2020ikq,%
Buonocore:2021rxx}.
In the large-$q_T$ region, where $q_T$ is of the order of the invariant
mass of the lepton pair $M$, the
fixed-order QCD corrections for the $q_T$ distribution are known
up to $\mathcal{O}(\alpha_S^2)$ in analytic
form\,\cite{Ellis:1981hk,Arnold:1988dp,Gonsalves:1989ar,Mirkes:1992hu,Mirkes:1994dp} and
up to $\mathcal{O}(\alpha_S^3)$ numerically
through the fully exclusive NNLO calculation of vector boson production
in association with jets\,\cite{Boughezal:2015dva,Ridder:2015dxa,Boughezal:2015ded,Boughezal:2016dtm,Gehrmann-DeRidder:2017mvr}.
However the bulk of the vector boson cross section lies in the small-$q_T$
region ($q_T\ll M$) where the reliability of the fixed-order expansion is
spoiled by the presence of large logarithmic corrections of the
type $\ln (M^2/q_T^2)$ due to the initial-state radiation of soft
and/or collinear partons.
In order to obtain reliable perturbative QCD predictions, 
the enhanced-logarithmic terms have to be evaluated and systematically
resummed to all orders in perturbation 
theory\,\cite{Dokshitzer:1978yd,Parisi:1979se,Collins:1984kg,Bozzi:2005wk,Catani:2010pd,Monni:2016ktx}.
Resummed calculations at different levels of theoretical accuracy
have been performed in Refs.\,\cite{Bozzi:2008bb
,Bozzi:2010xn,Banfi:2012du,Guzzi:2013aja,Catani:2015vma,Coradeschi:2017zzw,Bizon:2018foh,Bizon:2019zgf,%
Alioli:2021qbf}
also applying methods from Soft Collinear Effective
Theory\,\cite{Becher:2010tm
,Becher:2011xn,Ebert:2016gcn,Becher:2019bnm,Ebert:2020dfc,Becher:2020ugp,
Billis:2021ecs}
and transverse-mo\-men\-tum dependent
factorisation \cite{Collins:2011zzd
,Collins:2012uy,Collins:2014jpa,Scimemi:2017etj,Bertone:2019nxa,
Bacchetta:2019sam}.

In this Letter we apply the QCD transverse-momentum resummation formalism of Refs.\,\cite{Bozzi:2005wk,Bozzi:2010xn,Catani:2015vma}
for the case of $Z/\gamma^*$ boson production up to N$^3$LL accuracy.
We analytically include all the N$^3$LO terms at small-$q_T$ reaching full
N$^3$LL+N$^3$LO accuracy in the small-$q_T$ region\,\footnote{Sometimes in the literature
  this is referred as N$^3$LL$'$ accuracy.}.
We implement our resummed calculation in the public
numerical program {\ttfamily DYTurbo}\,\cite{Camarda:2019zyx}  which provides fast and
numerically precise predictions
both for resummed and fixed-order QCD calculations
including the full kinematical dependence of the decaying lepton pair
with the corresponding spin correlations and the finite
value of the $Z$ boson width. 
We consistently match our resummed predictions with the NNLO numerical results
at large-$q_T$ calculated in Refs.\,\cite{Ridder:2015dxa,Gehrmann-DeRidder:2017mvr}
and reported in Ref.\,\cite{Bizon:2019zgf}
thus including the $\mathcal{O}(\alpha_S^3)$ corrections for the entire
spectrum of $q_T$.
By using the connection between the $q_T$ resummation and the
$q_T$ subtraction formalism\,\cite{Catani:2007vq} for fixed-order calculations
we analytically\,\cite{Ablinger:2018sat} expanded the resummed results
thus providing predictions for
fiducial cross section of the Drell--Yan process
both at N$^3$LL+N$^3$LO and at N$^3$LO which, to our
knowledge, have never appeared in the literature. 
Higher-order calculations beyond NLO QCD are definitely an hard task and
are based on forefront and highly-specialized computations and numerical codes.
The calculation presented in this paper is released as public software\,\cite{dyturbo}
with the aim of facilitating an efficient and wide spread of the results to the theoretical and experimental communities.

We briefly review the resummation formalism
developed in Refs.\,\cite{Bozzi:2005wk,Bozzi:2010xn,Catani:2015vma} highlighting the main
aspect relevant for our calculation.
We consider the process
\beqn
\label{eq1}
h_1 + h_2 \to V +X \to l_3+l_4+X,
\eeqn
where $V$ denotes the vector boson\,\footnote{In this paper we
explicitly consider the case $V=Z/\gamma^*$, however our analytic results
and the ensuing numerical implementation can be 
extended for the
generic case of the production of colourless high-mass systems.}
produced
by the colliding hadrons $h_1$ and $h_2$ with a centre--of--mass energy $s$,
while $l_3$ and $l_4$ are the final state leptons produced by the $V$ decay.

The hadronic
cross section,
fully differential in the lepton kinematics,
is completely specified in terms of  the transverse-momentum ${\bf q_T}$
(with $q_T=\sqrt{{\bf q_T}^2}$), the rapidity $y$
and the invariant
mass $M$ of the lepton pair, and by two additional variables ${\bf \Omega}$
that specify the angular distribution of the leptons
with respect to the vector boson momentum.
The differential hadronic cross section can be written as
\beqn
\label{diffXS}
    \frac{d {\sigma}_{h_1h_2\to l_3l_4}}{d^2{\bf q_T}dM^2dy d\bf{\Omega}}
    ({\bf q_T},M^2,y, s, {\bf\Omega})
&=&
\sum_{a_1,a_2}\int_0^1dx_1\int_0^1dx_2 
\,f_{a_1/h_1}( x_1,\mu_F^2)\,f_{a_2/h_2}(x_2,\mu_F^2) \nonumber \\
&\times&\,
{\frac{d{\hat\sigma}_{a_1a_2\to l_3l_4}}{d^2{\bf q_T}\,d{M^2}\,d\hat y\,d{\bf \Omega}}} 
({\bf q_T},M,\hat y,\hat s,{\bf \Omega};\alpha_S,\mu_R^2,\mu_F^2)\,,
    \eeqn
    where $f_{a/h}(x,\mu_F^2)$ ($a=q_f,\bar q_f,g$) are the parton densities
    of the colliding hadron $h$, 
$\hat s = x_1 x_2 s$ is
the square of the partonic  centre--of--mass energy, 
$\hat y=y-\ln\sqrt{x_1/x_2}$ is the vector boson rapidity with respect to the
colliding partons,
$\mu_R$ and  $\mu_F$
are the renormalization and factorization scales. 
The last factor in the right-hand side of Eq.\,(\ref{diffXS})  is
multi-differential partonic cross sections,
computable in perturbative QCD as a series expansion in the strong coupling $\alpha_S=\alpha_S(\mu_R)$,
which will be denoted in the following  by the shorthand notation $[d \hat\sigma_{a_1a_2\to l_3l_4}]$.

The partonic cross section can be decomposed as
\beqn
\label{partXS2}
    [d \hat\sigma_{a_1a_2\to l_3l_4}]= [d \hat\sigma^{({\rm res.})}_{a_1a_2\to l_3l_4}]+
    [d \hat\sigma^{({\rm fin.})}_{a_1a_2\to l_3l_4}]
\eeqn
where the
first term on the right-hand side of Eq.\,(\ref{partXS2})
is the resummed component
which contains all the logarithmically-enhanced contributions of the type
$\alpha_S^n\,M^2/q_T^2\ln^m(M^2/q_T^2)$ (with $0\leq m \leq 2n-1$)
that have to be resummed to all orders,
while the second term is the finite component which
can be evaluated at fixed order in perturbation theory.

We perform the resummation in the impact-parameter space $b$\,\cite{Parisi:1979se}. The resummed 
component can then be written as 
\beqn
\label{resum}
\left[ {d{\hat \sigma}_{a_1a_2\to l_3l_4}^{(\rm res.)}} \right]
= \sum_{b_1,b_2=q,{\bar q}} \!\!
\frac{{d{\hat \sigma}^{(0)}_{b_1b_2\to l_3l_4}}}{d{\bf{\Omega}}} \;
\frac{1}{\hat s} \;
\int_0^\infty \frac{db}{2\pi} \; b \,J_0(b q_T) 
\;{\cal W}_{a_1a_2,b_1b_2\to V}(b,M,\hat y,\hat s;\alpha_S,\mu_R^2,\mu_F^2) \;,
\eeqn
where
$J_0(x)$ is the $0$th-order Bessel function,
the factor $d{\hat \sigma}^{(0)}_{b_1b_2\to l_3l_4}$ 
is the Born level differential cross section for the partonic subprocess
$q\bar q\to V \to l_3l_4$.

The function
${\cal W}_{V}(b,M,\hat y,\hat s)$ 
can be expressed in an 
exponential form
by considering 
the `double' $(N_1,N_2)$ Mellin moments 
with respect to the variables 
$z_1=e^{+\hat y}M/{\sqrt{\hat s}}$ and $z_2=e^{-\hat y}M/{\sqrt{\hat s}}$ 
at fixed $M$\,\footnote{
For the sake of simplicity
the explicit dependence  on parton indices (which are relevant for the exponentiation in the multiflavour space)
and the Mellin indices are understood.
The interested reader can find the details in Ref.\,\cite{Bozzi:2005wk} (in particular  Appendix A) and Ref.\cite{Bozzi:2007pn}.}
\cite{Bozzi:2005wk,Bozzi:2007pn}
\begin{equation}
\label{wtilde}
{\cal W}_{V}(b,M;\alpha_S,\mu_R^2,\mu_F^2)
={\cal H}_V\left(M; 
\alpha_S,M/\mu_R,M/\mu_F,M/Q \right) 
\times \exp\{{\cal G}(\alpha_S,\tL;M/\mu_R,M/Q)\}\,,
\end{equation}
where 
we have introduced
the logarithmic expansion parameter $\tL\equiv \ln ({Q^2 b^2}/{b_0^2}+1)$
with $b_0=2e^{-\gamma_E}$ ($\gamma_E=0.5772...$ 
is the Euler number).
The scale $Q\sim M$ 
is the resummation scale \cite{Bozzi:2003jy}, 
which parameterizes the
arbitrariness in the resummation procedure.

The process dependent function ${\cal H}_V$\,\cite{Catani:2000vq,Catani:2013tia}
includes
the hard-collinear contributions 
and it can be expanded in powers of $\alpha_S$ as
\begin{equation}
\label{hexpan}
{\cal H}_V(M;\alpha_S)=
1+ \frac{\alpha_S}{\pi} \,{\cal H}_V^{(1)} 
+ \left(\frac{\alpha_S}{\pi}\right)^2 
\,{\cal H}_V^{(2)}
+ \left(\frac{\alpha_S}{\pi}\right)^3 
\,{\cal H}_V^{(3)}+\dots \;.
\end{equation}

The universal (process independent) form factor $\exp\{{\cal G}\}$ 
in the right-hand side of Eq.~(\ref{wtilde})
contains all
the terms that order-by-order in $\alpha_S$ are logarithmically divergent 
as $b \to \infty$ (i.e.\ $q_T\to 0$). 
The resummed logarithmic expansion of ${\cal G}$ reads
\begin{equation}
\label{exponent}
{\cal G}(\alpha_S,\tL)=\tL\, g^{(1)}(\alpha_S \tL)
+g^{(2)}(\alpha_S \tL) 
+\frac{\alpha_S}{\pi} \;g^{(3)}(\alpha_S \tL)+
\left(\frac{\alpha_S}{\pi}\right)^2 \;g^{(4)}(\alpha_S \tL)+\dots\,,
\end{equation}
where the functions $g^{(n)}$ control and resum the $\alpha_S^k\tL^{k}$ (with $k\geq 1$) logarithmic terms in the exponent of Eq.\,(\ref{wtilde})
due to soft and collinear radiation. 
At NLL+NLO we include the functions $g^{(1)}$, $g^{(2)}$ and ${\cal H}_V^{(1)}$, 
at NNLL+NNLO we also include the functions $g^{(3)}$ and ${\cal H}_V^{(2)}$\,\cite{Catani:2012qa,Gehrmann:2012ze}. In order to reach full
N$^3$LL+N$^3$LO accuracy in the small-$q_T$ region
(i.e.\ including {\itshape all} the $\mathcal{O}(\alpha_S^3)$ terms) we have
included
the functions $g^{(4)}$\cite{Li:2016ctv,Henn:2019swt,vonManteuffel:2020vjv} and ${\cal H}_V^{(3)}$.
The function ${\cal H}_V^{(3)}$ has been determined by exploiting
its relation with the matching coefficients of the transverse-momentum dependent
parton densities (TMD) calculated in Refs.\,\cite{Luo:2019szz,Ebert:2020yqt} (see also Refs.\,\cite{Baikov:2009bg,Lee:2010cga,Gehrmann:2010ue}).
The Mellin moments of the function ${\cal H}_V$ have been calculated using the method of Ref.\,\cite{Albino:2009ci}, and the {\tt FORM}\,\cite{Kuipers:2012rf}
packages {\tt summer}\,\cite{Vermaseren:1998uu} and {\tt harmpol}\,\cite{Remiddi:1999ew}.
The evolution of parton densities in Mellin space, and the Mellin moments of the splitting functions are calculated with the package
{\tt QCD-PEGASUS}\,\cite{Vogt:2004ns}, the Mellin inversion and the Fourier--Bessel inverse transform from the impact-parameter space are
performed numerically as discussed in Ref.\,\cite{Camarda:2019zyx}.

The function ${\cal G}$ is singular when $\alpha_S\tL=\pi/\beta_0$ (where $\beta_0$ is the one-loop coefficient of the QCD $\beta$ function)
which corresponds to the region of transverse-momenta
of the order of the scale of the Landau pole of the QCD coupling
or $b^{-1}\sim \Lambda_{QCD}$. This signals
that a truly non-perturbative (NP) region is approached and 
perturbative results (including resummed ones) are not reliable. In this region
a model for NP QCD effects,
which has to include a regularization of the singularity of
the function ${\cal G}$, is necessary. In our calculation we explicitly implemented 
the so-called {\itshape Minimal Prescription}\,\cite{Catani:1996yz,Laenen:2000de,Kulesza:2002rh} which has the advantage to regularize the Landau singularity
in resummed calculations {\itshape without}
introducing {\itshape higher-twist} power-suppressed contributions of the type
$\mathcal{O}(\Lambda_{QCD}/Q)$.
Power-suppressed contributions can certainly be relevant
at very small transverse-momentum ($q_T\sim \Lambda_{QCD}$) and should
be eventually included, taking into account the delicate interplay with the leading-twist term,
in order to correctly describe the experimental data in that region.
In this Letter we have included the NP contribution in the form of a NP form factor $S_{NP}=\exp\{-g_{NP}\,b^2\}$ with $g_{NP} = 0.6$~GeV$^2$\,\footnote{This
value is of the same order of the ones typically fitted in the literature, see e.g.\ Refs.\,\cite{Kulesza:2002rh,Konychev:2005iy,Guzzi:2013aja}.}
which multiplies the perturbative form factor $\exp\{{\cal G}(\alpha_S,\tL)\}$,
leaving a more detailed analysis of the inclusion of the power-suppressed NP contribution to future work.

Finally the finite component $d \sigma^{({\rm fin.})}_{a_1a_2\to l_3l_4}$ has to be evaluated starting from the usual fixed-order
perturbative truncation of the partonic cross section and subtracting the expansion of the
resummed part at the same perturbative order (see Eq.\,(\ref{partXS2})):
\begin{equation}
\label{fincomp}
\Big[d{\hat \sigma^{({\rm fin.})}_{a_1a_2\to l_3l_4}}\Big]=
\Big[d{\hat \sigma_{a_1a_2\to l_3l_4}}\Big]_{\rm f.o.}
\,-\,
\Big[d{\hat \sigma^{({\rm res.})}_{a_1a_2\to l_3l_4}}\Big]_{\rm f.o.}\,.
\end{equation}
We have performed the analytic expansion of the resummed component Eq.\,(\ref{resum})
up to
$\mathcal{O}(\alpha_S^3)$ while the fixed-order cross section at large $q_T$ (formally at $q_T>0$) can
be obtained
from the the fully-exclusive  
computation of vector boson production 
in association with a jet at LO, NLO\,\cite{Campbell:2010ff}
and NNLO\,\cite{Boughezal:2015dva,Ridder:2015dxa,Boughezal:2015ded,Boughezal:2016dtm,Gehrmann-DeRidder:2017mvr}. 
We observe that both the fixed-order cross section and the expansion of the resummed part are separately divergent with the same small-$q_T$ limit
and the finite component formally satisfies the equation\,\cite{Camarda:2019zyx}
\begin{equation}
\label{fincomp2}
\lim_{q_T \to 0} \, q_T \,
d{\sigma^{({\rm fin.})}_{h_1h_2\to l_3l_4}}
= 0 \,. 
\end{equation}
We have checked that our 
 analytic expression for the expansion of the resummed part agrees in the small-$q_T$ limit with the NNLO fixed-order results  reported in Ref.\,\cite{Bizon:2019zgf}
 at permille level down to
 $q_T\sim 4$~GeV\,\footnote{Below the threshold of
   $q_T\sim 4$~GeV the agreement of the $\mathcal{O}(\alpha_S^3)$ corrections  
   deteriorates (this is not the case for the $\mathcal{O}(\alpha_S)$ and $\mathcal{O}(\alpha_S^2)$ results).
   Validation with higher-statistic results would be necessary in this region.}.

In the following we consider  $Z/\gamma^*$ production and leptonic decay at the LHC. 
We present resummed predictions at NLL+NLO, NNLL+NNLO and N$^3$LL+N$^3$LO accuracy,
matching our computation with the fixed-order results at large-$q_T$
respectively at LO, NLO and NNLO.
The hadronic cross section is obtained by convoluting
the partonic cross section in Eq.\,(\ref{partXS2}) with the parton densities functions
(PDFs) from the NNPDF3.1 set\,\cite{Ball:2017nwa} at
NNLO  with $\as(m_Z^2)=0.118$
where we have evaluated $\as(\mu_R^2)$ at $(n\!+\!1)$-loop order
at N$^n$LL+N$^n$LO accuracy.
In the case of $Z$ production, because of the axial coupling,  additional Feynman diagrams with quark loops contribute
to the cross-section at $\mathcal{O}(\alpha_S^2)$ and $\mathcal{O}(\alpha_S^3)$. 
Their contribution cancels out for each isospin multiplet when massless quarks are considered. The
effect of a finite top-quark mass in the third generation has been considered and found extremely
small at $\mathcal{O}(\alpha_S^2)$\,\cite{Dicus:1985wx,Gonsalves:1989ar} while the finite mass top-quark contribution at $\mathcal{O}(\alpha_S^3)$
remains to be derived\,\cite{Gehrmann:2021ahy}. Therefore these contributions have currently been neglected in our calculation.
We use the so called $G_\mu$ scheme for EW couplings
with input parameters 
$G_F = 1.1663787\times 10^{-5}$~GeV$^{-2}$,
$m_Z = 91.1876$~GeV, $\Gamma_Z=2.4952$~GeV, $m_W = 80.379$~GeV.
Our calculation implements 
the leptonic decays $Z/\gamma^* \to l^+l^-$ 
and we include the effects of the $Z/\gamma^*$ interference and of the
finite width $\Gamma_Z$
of the 
$Z$ boson
with the corresponding spin correlations and the full dependence  
on the kinematical variables of final state leptons.
This allows us to take into account the typical 
kinematical cuts on final state leptons that are
considered in the experimental analysis. 
The resummed calculation  at fixed lepton momenta requires a $q_T$-recoil procedure.
We implement the general procedure described in Ref.\,\cite{Catani:2015vma}
which is equivalent to compute the Born level distribution 
$d{\sigma}^{(0)}$ of Eq.\,(\ref{resum})
in the Collins--Soper rest frame\,\cite{Collins:1977iv}.

\begin{figure}[t]
\begin{center}
  \includegraphics[width=0.7\textwidth]{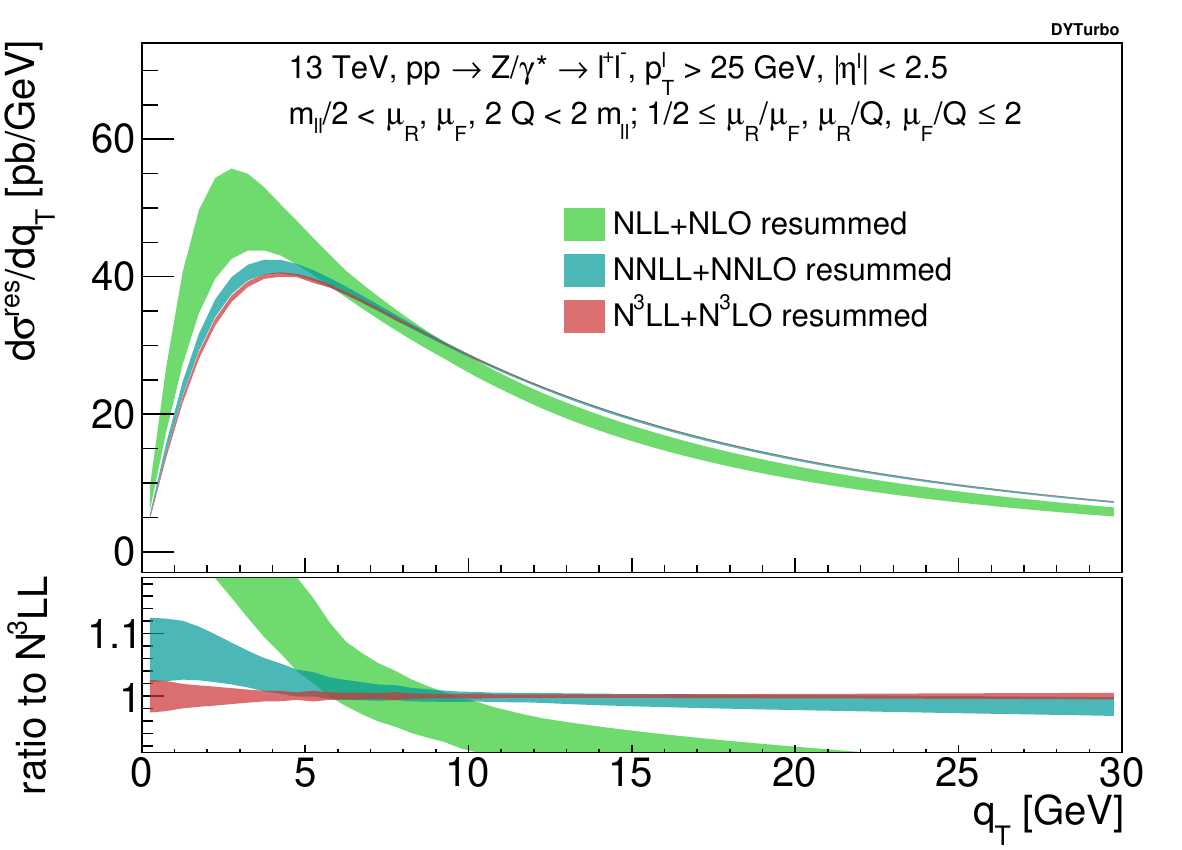}
\end{center}
\caption{
  \label{fig1}
  {\em
    The $q_T$ spectrum of $Z/\gamma^*$ bosons with lepton selection cuts at  the LHC ($\sqrt{s}=13$~TeV)
    at various perturbative orders.
    Resummed component (see Eq.\,(\ref{partXS2})) of the hadronic cross-section
with scale variation bands as  defined in the text. 
}}
\end{figure}

\begin{figure}[t]
\begin{center}
  \includegraphics[width=0.7\textwidth]{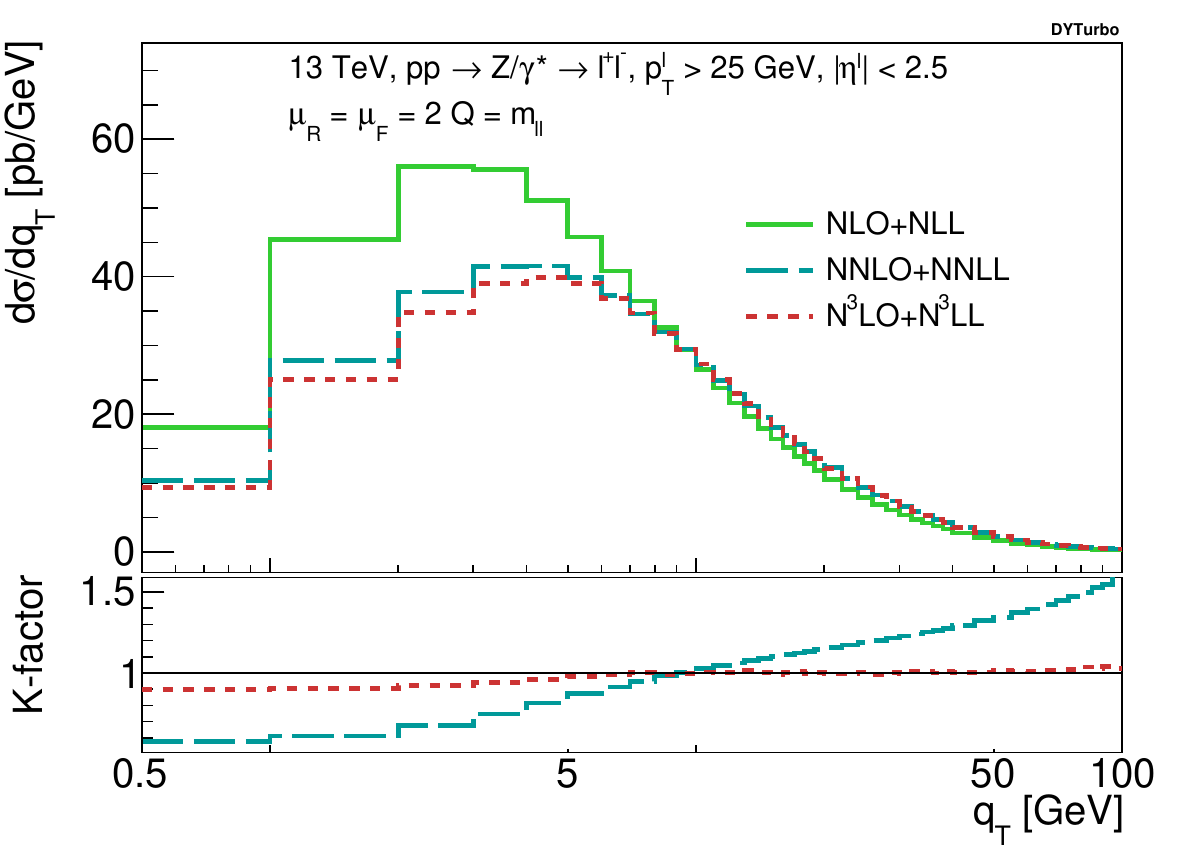}
\end{center}
\caption{
  \label{fig2}
  {\em
    The $q_T$ spectrum of $Z/\gamma^*$ bosons with lepton selection cuts at  the LHC ($\sqrt{s}=13$~TeV)
    at various perturbative orders.
    Full matched results between resummed and finite part of the hadronic cross section 
at central values of the scales.
}}
\end{figure}

We have applied the resummation formalism to the production of
$l^+ l^-$ pairs from  $Z/\gamma^*$ decay at
the LHC ($\sqrt{s}=13\,$TeV) with the following
fiducial cuts:
the leptons are required to have transverse momentum $p_T>25\,$GeV,
pseudo-rapidity $|\eta|<2.5$ while the lepton pair system, is required
to have  invariant mass $66<M<116\,$GeV and transverse momentum
$q_T<100$~GeV\,\footnote{In order to match with the
NNLO numerical results at large-$q_T$
we follow the kinematical selection cuts applied in Ref.\,\cite{Bizon:2019zgf}.}.

In Fig.\,\ref{fig1} we show the
resummed component (see Eq.\,(\ref{fincomp2})) of the transverse-momentum
distribution in the small-$q_T$ region. In order
to estimate the size of yet uncalculated higher-order terms
and the ensuing perturbative uncertainties we present
the  dependence of the resummed component on the
auxiliary scales $\mu_F$, $\mu_R$ and
$Q$.
The scale dependence band 
is obtained 
through independent variations of 
$\mu_F$, $\mu_R$ and $Q$
in the range
$M/2 \leq \{ \mu_F, \mu_R, 2Q \} \leq 2M$ with the constraints
$0.5\leq \{ \mu_F/\mu_R, Q/\mu_R,
Q/\mu_F \}\leq 2$\,\footnote{In order to estimate the
  $Q$ scale dependence of the
  resummed component we set the logarithmic expansion
  parameter to be $L = \ln ({Q^2 b^2}/{b_0^2})$ which is equivalent to $\tL$
  in the small-$q_T$ region.}. The lower panel shows the ratio of the distribution with respect to the N$^3$LL+N$^3$LO prediction
at the central value of the scales $\mu_F=\mu_R=2Q= M$.
We observe that the NLL+NLO and NNLL+NNLO scale dependence bands do not
overlap thus showing that the  NLL+NLO scale variation
underestimates the true perturbative uncertainty.
This feature was observed and discussed in Ref.\cite{Catani:2015vma}.
Conversely the NNLL+NNLO and N$^3$LL+N$^3$LO scale variation bands
do overlap in the entire region $q_T<30\,$GeV
(except that they nearly overlap in the window $1<q_T<4\,$GeV) thus
suggesting that, from NNLL+NLO, missing higher order corrections are correctly estimated by scale variations.
We also observe that the scale dependence is reduced by a factor
of 2 (or more) going from NNLL+NNLO to  N$^3$LL+N$^3$LO:
the scale variation at  N$^3$LL+N$^3$LO (NNLL+NNLO) is around $\pm 0.8\%$ ($\pm 2.5\%$) at the peak ($q_T\sim 4\,$GeV),
then it reduces at $\pm 0.3\%$ ($\pm 0.8\%$) level at $q_T\sim 12\,$GeV and increase up to $\pm 0.4\%$ ($\pm 1.4\%$) level at $q_T\sim 25\,$GeV. 
Finally, we note that in the low $q_T$ region non-perturbative effects are expected to become important. In particular
by considering variations of the NP parameter in the range $0.3 \leq g_{NP} \leq 0.9$~GeV$^2$ we obtain the following additional uncertainties for the N$^3$LL+N$^3$LO resummed prediction in
Fig.\,\ref{fig1}: $\pm 0.3\%$  at $q_T\sim 25\,$GeV,  $\pm 0.6\%$  at $q_T\sim 12\,$GeV and $\pm 0.7\%$  at $q_T\sim 4\,$GeV. For $q_T\lesssim 4\,$GeV the NP uncertainties rapidly
increase at few percent level. These NP uncertainties have a delicate interplay with the non-perturbative parton densities uncertainties which deserve a careful analysis.

In Fig.\,\ref{fig2} we show the 
resummed $q_T$ distribution matched with the finite part 
at LO, NLO and NNLO. 
The auxiliary scales have been fixed to their central values
$\mu_F=\mu_R=2Q= M$\,\footnote{
  Central scales for the fixed-order result
  have been set to $\mu_F=\mu_R= \sqrt{M^2+q_T^2}$.
  The calculation of the scale variation band of the matched distribution would require the knowledge
  of the NNLO fixed-order result at $q_T>0$ for different values of $\mu_F$ and $\mu_R$.}.
The lower panel shows the $K$-factors $K_{N^nLO}$  defined as the ratio of between the N$^n$LL+N$^n$LO and the 
N$^{n-1}$LL+N$^{n-1}$LO predictions (with $n=2,3$).  By looking at the $K$-factors we observe that the impact of the N$^3$LL+N$^3$LO (NNLL+NNLO)
corrections with respect to the previous order is around $-4\%$ ($-19\%$) at the peak, 
then it becomes $-0.1\%$ ($+22\%$) at $q_T\sim 30\,$GeV and increase to $+3\%$ ($+55\%$)  at $q_T\sim 90\,$GeV.   

\begin{figure}[t]
\begin{center}
  \includegraphics[width=0.7\textwidth]{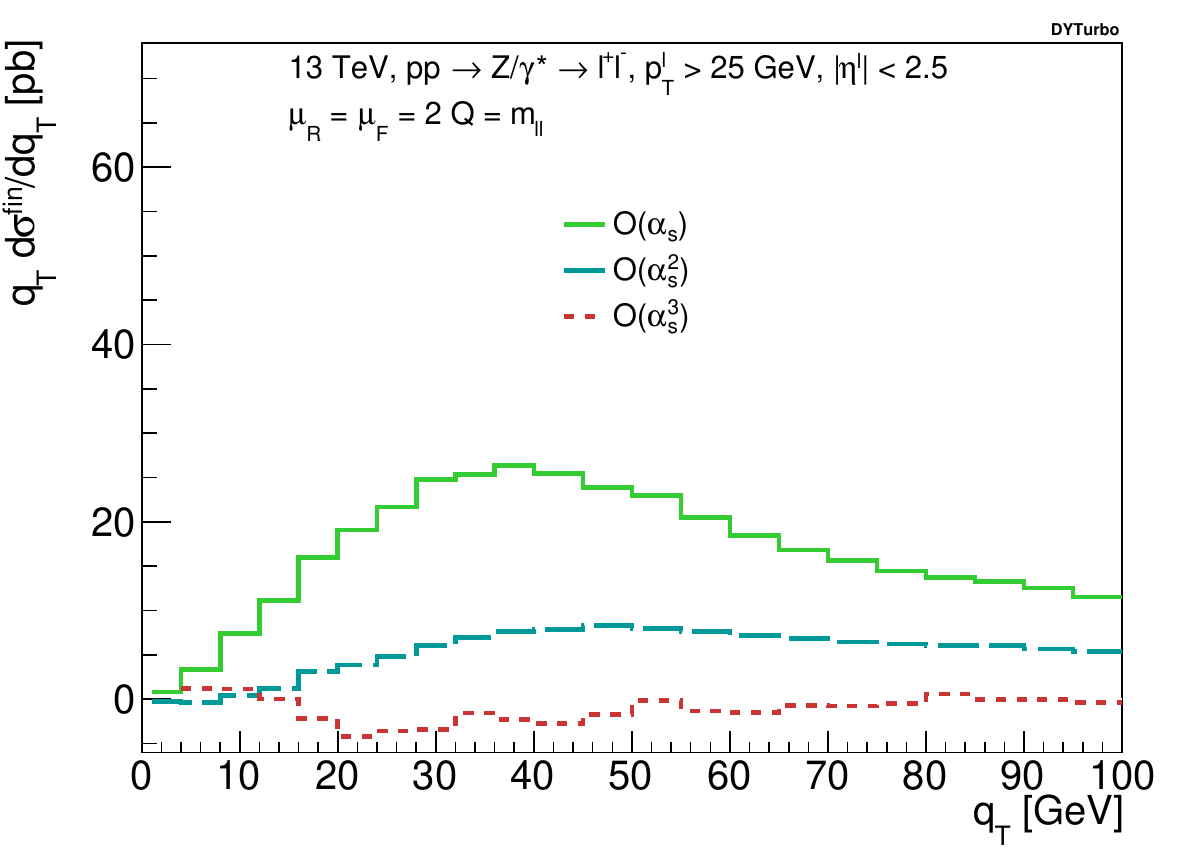}
\end{center}
\caption{
  \label{fig3}
  {\em
    The $q_T$ spectrum of $Z/\gamma^*$ bosons with lepton selection cuts at  the LHC ($\sqrt{s}=13$~TeV).
   Contribution of the finite part of the hadronic cross section
for central values of the scales at $\mathcal{O}(\alpha_S)$, $\mathcal{O}(\alpha_S^2)$ and $\mathcal{O}(\alpha_S^3)$ (see Eq.\,\ref{fincomp2}). 
}}
\end{figure}

\begin{table}[!ht]
  \begin{center}
        \begin{tabular}{lccc}
      \toprule
Order            					&  NLO   	& NNLO       	&  N$^3$LO   	 \\
\midrule                                                                                                                         
$\sigma(pp\to Z/\gamma^* \to l^+l^-)$ [pb]&  $766.3 \pm 1 $ &  $757.4 \pm 2 $ &  $746.1 \pm 2.5$   \\
\midrule                                                                                                                         
\midrule                                                                                                                         
Order            					& NLL+NLO & NNLL+NNLO & N$^3$LL+N$^3$LO   	 \\
\midrule                                                                                                                         
$\sigma(pp\to Z/\gamma^* \to l^+l^-)$ [pb]& $ 773.7 \pm 1 $ & $759.8  \pm 2 $ & 749.6 $\pm 2.5$   \\ 
\bottomrule
    \end{tabular}
  \end{center}
\caption{{\em Fiducial cross sections at the LHC ($\sqrt{s}=13$~TeV): fixed-order
    results and corresponding resummation results obtained with the {\tt DYTurbo} numerical program. The uncertainties
    on the values of the cross sections plots refer
    to an estimate of the numerical uncertainties in the integration.
}}
\label{table}
\end{table}

By exploiting the connection between the $q_T$ resummation and the $q_T$ subtraction formalism\,\cite{Catani:2007vq}
we are able to  provide fixed-order results for
fiducial cross sections up to N$^3$LO\,\footnote{A fully consistent N$^3$LO calculation for hadronic cross sections would require PDFs at the corresponding order which are currently not
  available. Uncertainties from missing higher order PDFs have been studied in Refs.\,\cite{Forte:2013mda,Moch:2017uml,Moch:2018wjh}.}. 
In Table~\ref{table} we report the predictions for the cross section
in the fiducial region at NLO, NNLO and N$^3$LO, NLL+NLO, NNLL+NNLO,  N$^3$LL+N$^3$LO
fixing the auxiliary scales to their central values.
The N$^3$LO corrections decrease the NNLO cross-section at $-1.5\%$ level and the resummation effects further enhance the N$^3$LO  result by $+0.5\%$.
We observe that the $K$-factor between the N$^3$LO and NNLO results is $0.985$ which is comparable with results reported in Table I of Refs.\,\cite{Duhr:2020seh,Duhr:2020sdp}.

Generally speaking  scale dependence cannot be regarded as a consistent estimate of the perturbative uncertainty
because the effect due to uncalculated higher-order terms is typically larger than conventional scale dependence (see e.g.\ the comments
on the scale variation bands of Fig.\,\ref{fig1}).
A more realistic uncertainty estimate of the ``true'' perturbative uncertainty  can be obtained, for instance,
by comparing two subsequent orders of the expansion at central values of the scales and using half of the difference between them to
assign the perturbative uncertainty\,\cite{Catani:2018krb}. This procedure leads (see Table~\ref{table}) to an
uncertainty of about $\pm 2\%$ ($\pm 4\%$) at NLO (NLL+NLO), $\pm 0.7\%$ ($\pm 1\%$) at NNLO (NNLL+NNLO) and $\pm 0.8\%$ ($\pm 0.7\%$) at N$^3$LO (N$^3$LL+N$^3$LO).
As expected (see comments below on fiducial cuts) this procedure shows  a better convergence on the uncertainties of the resummed perturbative expansion.

In order to judge the numerical stability of the matching procedure and
the effects of the power-suppressed terms we consider the 
contrubution of the finite part of the cross section.
We show in Fig.\,\ref{fig3} the finite part of the cross section  
for central values of the scales at $\mathcal{O}(\alpha_S)$, $\mathcal{O}(\alpha_S^2)$ and $\mathcal{O}(\alpha_S^3)$.
The effect of the finite component smoothly vanish as $q_T \to 0$ (see Eq.\,\ref{fincomp2}) and gives a small contribution to the matched result in the small $q_T$ region:
the integral over the ranges $4<q_T<20\,$GeV
and $1<q_T<4\,$GeV of the LO
finite component represents respectively the $1.5\%$ and $0.12\%$ of the
NLL+NLO fiducial cross section in Table~\ref{table},
the $\mathcal{O}(\alpha_S^2)$ correction in the same ranges is
respectively the $0.10\%$ and $-0.04\%$ of the NNLL+NNLO result
while the $\mathcal{O}(\alpha_S^3)$ correction
in the range $4<q_T<20\,$GeV the $0.16\%$ of the N$^3$LL+N$^3$LO.
As previously observed below  $q_T\sim 4$~GeV the agreement with the $\mathcal{O}(\alpha_S^3)$ results of Ref.\,[67] (see Fig.\,\ref{fig2})
deteriorates. However the finite component gives a tiny contribution in the small $q_T$ region and we
thus avoided to include in our results the finite part at $\mathcal{O}(\alpha_S^3)$ for $q_T<4$ GeV.

The results in Table~\ref{table} have been obtained applying the
symmetric lepton $p_T$ cuts previously defined. It is well
known\,\cite{Catani:1997xc,Catani:2018krb} that in the case of
symmetric cuts fixed-order calculations are
affected by perturbative (soft-gluon) instabilities at higher orders.
The results in Table~\ref{table} are obtained with a lower integration
limit for the finite part of the cross section fixed to $q_{T
  {cut}}=0.5$\,GeV and the numerical uncertainties include an estimate
of the corresponding systematic uncertainty.  More accurate
fixed-order results and an estimate of such uncertainty can be
obtained by evaluating the $q_{T {cut}}\to 0$ extrapolation or by a direct
calculation of perturbative power corrections of the type $\mathcal{O}((q_{T {cut}}/M)^p)$
with $p>0$ which are neglected for $q_{T {cut}}> 0$\,\cite{Ebert:2018gsn,Cieri:2019tfv,Ebert:2019zkb,Buonocore:2019puv,Ebert:2020dfc,Oleari:2020wvt}.  We stress however
that the inclusion of such contributions cannot improve the physical
predictivity of the fixed-order results in case of symmetric cuts which are affected by
sizable theoretical instabilities produced by the soft-gluon effects.

We have performed  the implementation of both the $q_T$ resummation
and $q_T$ subtraction formalism
for Drell--Yan processes up to N$^3$LL+N$^3$LO and N$^3$LO in the
{\tt DYTurbo} numerical program.
In this Letter we have illustrated the first numerical results
for the case of $Z/\gamma^*$ production and leptonic decay at the LHC.

\paragraph{Acknowledgments.} 
We gratefully acknowledge Stefano Catani for useful discussions and
comments on the manuscript
and Ludovica Aperio Bella for extensive tests of the numerical code.
This project has received funding from the European Union's Horizon 2020 research and innovation programme under
the Marie Sk\l odowska-Curie grant agreement number 754496 and under European Research Council grant agreement number 740006.

\bibliographystyle{utphys}
\bibliography{dyqtN3LL-fin}{}

\end{document}